# Mind-body interpretation of quantum mechanics[1]

Raoul Nakhmanson[2]

*The wave-particle duality is a mind-body one. In the real 3D-space there exists only the particle, the wave exists in its consciousness. If there are many particles, their distribution in accordance with the wave function represents a real wave in real space. Many worlds, Schrödinger cat, etc., exist only as mental constructions. The "waves of matter" are non-material. Feynman et al. taught quantum world "is like neither". Alas, they forgot living matter.*

This series of conferences is devoted to mysteries, puzzles and paradoxes in quantum mechanics (QM), in this year especially to quantum interference phenomenon. Therefore I begin with an explanation of the so-called wave-particle duality, the wave function and its interference, and if the time allows I will explain other "mysteries" such as entanglement, quantum randomness and uncertainty, and show an experimental feasibility to go beyond the scope of QM.

Following tradition let us consider the two-slit experiment (Fig.1). Some electrons emitted from the source *S* reach the registration plane *R* because the screen *SC* has two slits. In the plane *R* the electrons are registered as particles, but their spatial distribution looks like an interference of two coherent waves cut out by the screen *SC* from the initial wave generated by the source *S*. To emphasize is that it is not a many-body effect: The pattern is the same in the case when the source *S* is so week that not more than one electron flies between *S* and *R*, and we must conclude there is a "self-interference" of electrons. Seemingly electrons (and other particles) present themselves as well as waves. Hence the notion "wave-particle duality".

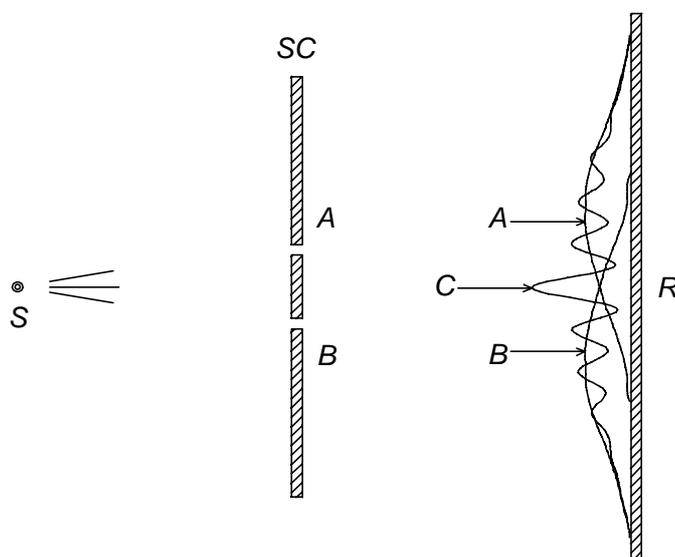

Fig. 1. Two-slit experiment. The distributions *A*, *B*, and *C* correspond to the cases when the slit *A* only, the slit *B* only, and the slits *A*, *B* both are opened, respectively.

To have the interference pattern we must accept that the "wave of matter" goes through both slits. But if we place two detectors behind the slits, only one of them registers the electron, hence it goes only through one slit. The other slit at the moment seemingly does not play any role and can be temporarily closed. If it is so, the two-slit experiment could be decomposed into two one-slit experiments which do not possess the interference being observed. Therefore we have an apparent conflict, and it is an essence of the "interference mystery".

---

[1] Shortened version of a report in IV Edition Workshop on MPP in QM, 31.08.2001, Gargnano, Italy.
[2] Nakhmanson@t-online.de

The attempts to solve this paradox went in two guidelines. The first, so-called "Copenhagen interpretation" of QM have reduced physical world to "observables". Wheeler declared: *"No elementary phenomenon is a phenomenon until it is a registered (observed) phenomenon."* [1]. In such a frame it is absurd to ask "which slit the electron has passed" if we look at interference. The problem with reduction of wave function had inspired here a many-world hypothesis. The Copenhagen interpretation is fruitless and the many-world nightmare cannot be accepted by common sense.

The second guideline includes different attempts to complete the mathematical formalism of QM with physical reality. Schrödinger tried to see a particle as a wave packet, de Broglie developed pilot-wave concept, Bohm attributed to each point of space a realistic quantum potential based on the wave function. All these attempts have no success: they don't predict something new hence cannot be proved experimentally, and they have inner problems, e.g. wave packets disperse, quantum potential has nonlocal links.

But the general and fatal mistake of these attempts is that they place a wave function into the real 3D-space. A wave function does not belong to physical reality. It is not in the real 3D-space, we cannot find and measure it there directly. If you allow me a pun, the **"waves of matter" are non-material.** Einstein justly called them "Gespensterfelder".

A wave function is a mental construction in an abstract configuration space. Nevertheless it controls the behavior of material objects. How is it possible? We don't know how. But we know a really working example: The human consciousness can control behavior of material objects (though we also do not know how). Following this analogy we must conclude that the particle has some kind of consciousness controlling its behavior, and the wave function is a product of this consciousness. The small size of an "elementary" particle must not confuse us: An object with a size of an elementary particle can accommodate much more Planck cells than not only the number of neurons in the human brain, but even than the total number of atoms of all known biological objects.

The wave-particle duality is a mind-body one. In the real 3D-space there exists only the particle, the wave exists in its consciousness. If there are many particles, their distribution in accordance with the wave function represents a real wave in real space. Many worlds, Schrödinger cat, Great Smoky Dragon, etc. exist only as virtual mental constructions.

The behavior of particles is purposeful, which is reflected in the teleological nature of physical laws (variation principles). Interacting particles exchange information. They need to have correlated notions about space and time, and in this sense one may speak of the preferred system like our Greenwich one. The "holism" of the Universe is informational in its nature. The "Internet" of matter exists probably from the time of Big Bang. Therefore the particles have a good impression of surroundings. One might expect that the civilization of particles has undergone a long evolution. It is possible that this civilization is past its prime already, and is now in the state of stagnation.

The wave function is the strategy of the particle. Receiving new information, the particle adjusts its strategy, that is, its wave function. This is the so-called "collapse" of the wave function. It occurs not in the real space, as is commonly thought, but in the consciousness of the particle, that is, locally and instantaneously on the common-sense scale. Contrary to the opinion of von Neumann, Wigner and others, in the general case the human consciousness has nothing to do with the collapse.

Such an interpretation of the wave function allows us to explain all QM mysteries, particularly interference. In the case of Fig. 1 the particle emitted by source $S$ in the direction of any slit goes through that slit. But its strategy takes into account a possibility to be emitted in the direction of another slit. These two possibilities interfere in the particle's consciousness and predict an oscillating distribution in the registration plane $R$. Following this instruction the particle interacts with the edges of the slit to have corresponding angles of deflection.

The wave function only sets the priorities. Taking them into account the particle makes a random choice. Such tactics leads to "impartial" exploration of all alternatives.

Because the least action principle can lead to the wave equation, we may expect interference of possibilities in purposeful behavior of living beings. Fifteen years ago I tried to perform a computer-simulated two-slit experiment with people (Fig. 2(*a*)). The "particle" comes from the source $S$ and moves up. The person under test can deflect it left and right, particularly can lead it through one of the slits in the screen $SC$ and further to the registration plane $R$. For comparison the program has also screens with only left or only right slit.

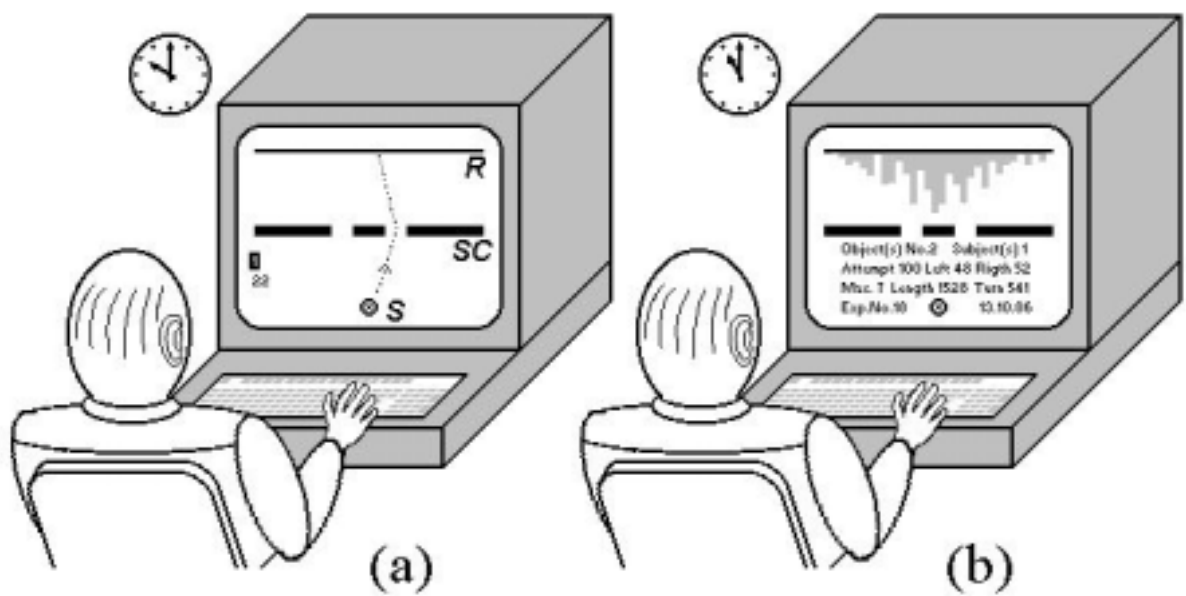

Fig. 2. Computer simulation of two-slit experiment with people.

The space between $S$ and $R$ was defined as a "forest". In the forest there are random distributed hidden "mushrooms" which develop themselves only if the particle touch them. The new mushrooms grow randomly over the forest while their quantity stays constant. The person was motivated to pick the mushrooms. After 100 to 200 attempts the distribution of events over the registration plane $R$ was displayed (Fig. 2(*b*)).

To get a stable result in such experiments one needs good statistics. It can be one person having stable characteristics and be tested repeatedly, or many persons having the same characteristics. Both requests correspond to monochromatic flow and are not easy to realize. Another problem is the density of mushrooms. Because the collision with a mushroom can be regarded as identification of the position of a particle and a possible change of its strategy, i.e. collapse of a wave function, such attempts were excluded from the statistics. Therefore for statistics it is better to have no mushrooms at all, that however destroys the motivation.

Although of the first results were promising, the problem with statistics was not solved then. The experiments were stopped and are continued only now.[3]

Let us come back to physics. Two or more particles may have a common strategy. In such a case their common wave function does not decompose into a product of partial wave functions. Such "entangled" particles, being separated in space, nevertheless act in a concerted way.

The information available to the particle is the knowledge of the past. For solving the variation problem, the particle must be able to predict what is in store for it. Prediction is a necessary attribute of any consciousness. A consciousness with the faculty of prediction is a non-mechanical hidden parameter that is possessed by the particles and that was overlooked by Bell in the formulation of his theorem.

---

[3] The modern version of the program can be sent per e-mail on request.

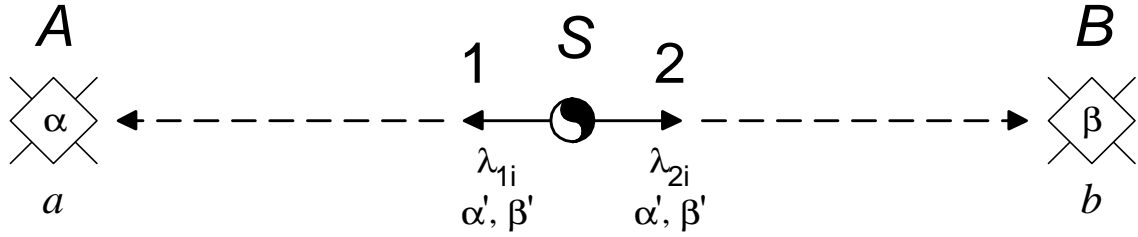

Fig. 3. Einstein-Podolsky-Rosen (EPR) experiment

Fig. 3 presents the scheme of Einstein-Podolsky-Rosen (EPR) experiment. The source *S* emits two entangled particles **1** and **2** possessing series of hidden parameters $\lambda_{1i}$ and $\lambda_{2i}$, respectively. The particles fly in opposite directions and are measured in distant points *A* and *B*. The measurers having adjusted conditions $\alpha$ and $\beta$, respectively, and give out results *a* and *b*.

The proof of Bell's theorem is based on the next assertion: If $P_a$ is a probability of result *a* measured on the particle **1** in the point *A* having a condition (e.g. angle of analyzer) $\alpha$, and $P_b$ is a probability of result *b* measured on the particle **2** in the distant point *B* having a condition $\beta$, then $\beta$ has no influence on the $P_a$, and vice versa. Herein Bell and others saw the indispensable requirement of local realism and "separability". Mathematically it can be written as

$$P_{ab}(\lambda_{1i},\lambda_{2i},\alpha,\beta) = P_a(\lambda_{1i},\alpha) \times P_b(\lambda_{2i},\beta) \quad \text{(Bell)}, \quad (1)$$

where $P_{ab}$ is the probability of the join result *ab*, and $\lambda_{1i}$ and $\lambda_{2i}$ are hidden parameters of particles **1** and **2** in an arbitrary local-realistic theory. Under the influence of Bell's theorem and the experiments following it and showing, that for entangled particles the condition (1) is no longer valid, some "realists" reject locality. In this case an instantaneous action at a distance is possible, and one can write

$$P_{ab}(\lambda_{1i},\lambda_{2i},\alpha,\beta) = P_a(\lambda_{1i},\alpha,\beta) \times P_b(\lambda_{2i},\beta,\alpha) \quad \text{(nonlocality)}. \quad (2)$$

In principle such a relation permits a description of any correlation between *a* and *b*, particularly predicted by QM and observed in experiments. But if the particle possesses consciousness, in the frame of local realism the condition (1) is not indispensable. Instead, one can write

$$P_{ab}(\lambda_{1i},\lambda_{2i},\alpha,\beta) = P_a(\lambda_{1i},\alpha,\beta') \times P_b(\lambda_{2i},\beta,\alpha') \quad \text{(prediction)}, \quad (3)$$

where $\alpha'$ and $\beta'$ are the conditions of measurements in points *A* and *B*, respectively, as they can be predicted by particles at the moment of their parting. If the prediction is good enough, i.e., $\alpha' \approx \alpha$ and $\beta' \approx \beta$, then (3) practically coincides with (2) and has all its advantages plus locality. The way for local-realistic description of nature is free.

There are two ways to confirm the consciousness of particles. The first is to prevent the particles from correct prediction the future, which should lead to nonstandard results. Such attempts were made by groups of Aspect, Alley, and Zeilinger [2-4]. The first group used periodically switching of analysers which was, of course, predictable. Two other groups used the "random" switching, but the randomness was borrowed from the object of study itself (i.e. the quantum world), which cannot be regarded as reasonable. Perhaps using a good human-programmed "quasi-random" generator is preferable.

The second way is to affect the particles with information. Let us illustrate this with some examples. In Fig. 4 an impulse of polarized light from source *S* is passed through a "black box" meaning that the observer-physicist knows what goes in and can measure what goes out, but does not know what is inside the box. In the case shown in Fig. 4(*a*) the light that comes out of the box can be either deflected to the right by a movable mirror *M* , or allowed to continue its way forward. Such control of the beam, somewhat like a railway switch, we call "force control".

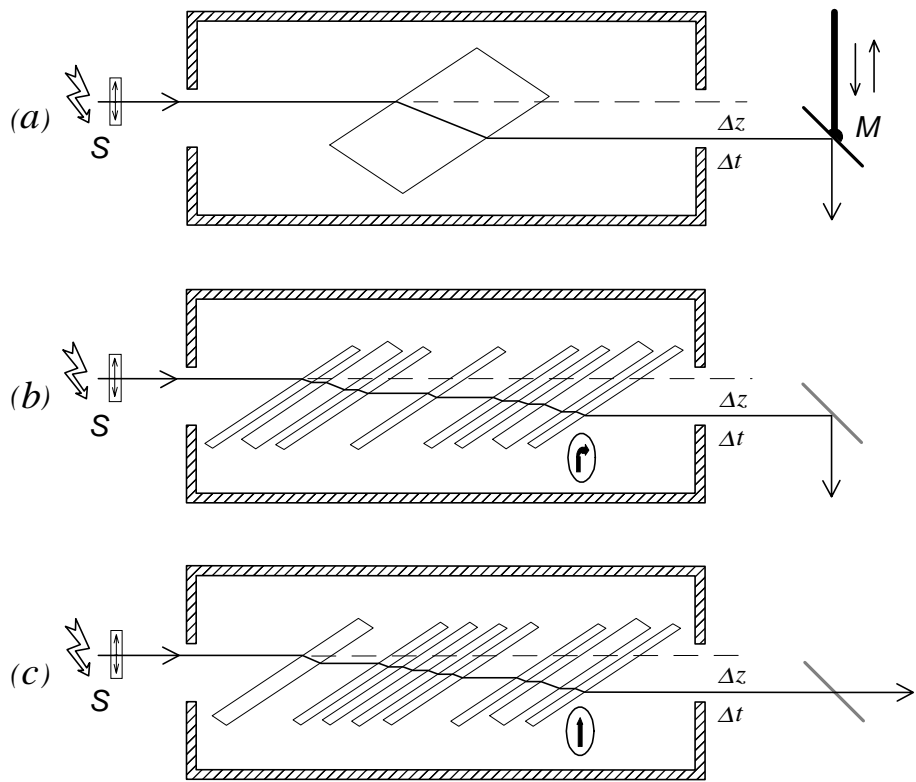

Fig. 4. (*a* ) - force control, (*b* ) and (*c* ) - informational control .

If we place inside the box a thick transparent glass plate on the path of the beam at the Brewster angle to the beam (as seen in Fig. 4(*a*)), we do not introduce any absorption or reflection, but the physicist can use his instruments to see that:

(1) (owing to the refraction in the plate) the beam going out of the box is parallel displaced to the right by a distance $\Delta z$ (as shown in Fig. 4(*a*));

(2) (owing to the fact that the speed of light in glass is slower than in air, and the path length is increased) the light will come out of the box with some delay $\Delta t$ .

And this is all that the physicist can find out without looking into the box.

In Figs. 4(*b*) and 4(*c*) in place of the total reflecting movable mirror *M* we have fixed semitransparent mirrors, and the thick glass plate in the box is split into eight thin plates of which two are thicker than the rest. Our physicist will not notice any of the changes made inside the box, because he will measure the same $\Delta z$ and $\Delta t$ . The photons, however, if they are intelligent and know English and Morse code, will read the following instructions:

$$\bullet - \bullet \quad \bullet \quad \bullet \bullet - \bullet \; = \text{REF (reflect)} \quad \text{in Fig. 4(}b\text{)} ,$$

$$- \quad \bullet \bullet \bullet \bullet \quad \bullet - \bullet \; = \text{THR (through)} \quad \text{in Fig. 4(}c\text{)} ,$$

and will carry them out by reflecting in Fig. 4(*b*) or passing through in Fig. 4(*c*). Such control of the beam, like the traffic signs at a crossroad, may be referred to as "informational control".

Note that carrying out of such "informational" experiments with elementary particles differs from anything that has been done in physics so far. And, of course, any consciousness can manifest itself only in a situation having alternative outcomes.

It is possible that the particles actually know all human languages and codes. But it would be safer to assume that we are dealing with a totally different civilization that knows nothing about us, so that our first contacts will run into difficulties. This problem is not new, and has been seriously considered in the framework of the Search for Extraterrestrial Intelligence project SETI. Its experts are inventing "cosmic" languages capable of developing communication from zero to a high semantic level. At the initial stage one could recommend trying such universal languages as mathematics and music. The starting point for identifying intelligence that may be much unlike ours, and for trying to establish contact with it, should be some very general property presumable inherent in any kind of intelligence. A good candidate for such a role is curiosity.

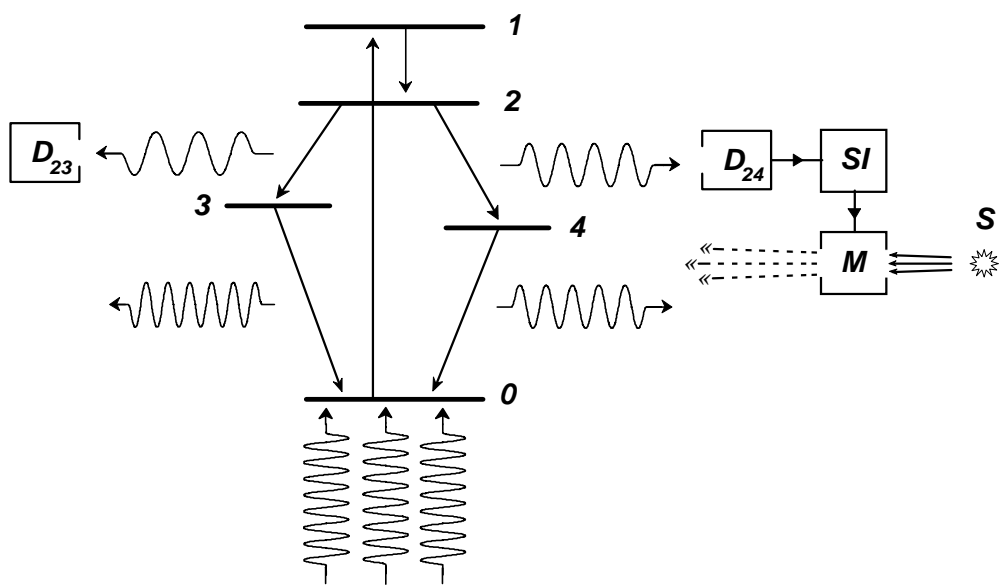

Fig. 5. Informational experiment with a single ion. 0, 1, 2, 3, and 4 are the energy levels; $D_{23}$ and $D_{24}$ are detectors, $S$ is the source of light, $M$ is the modulator, $SI$ is the source of information .

Figure 5 shows the scheme of an experiment that does not presume that the particles are aware of our culture in any way. Here a five-level quantum system, e.g., an individual ion in trap, with one low (0), one high (1), and three intermediate (2,3,4) energy levels is pumped by intensive radiation inducing the 0→1 transition, so that the ion does not stay in the state 0 but immediately is translated into the state 1 followed by metastable state 2. From it, the ion makes a spontaneous transition to the states 3 or 4, and later makes a transition to the state 0 completing the cycle. The radiation corresponding to some of the transitions 2→3, 2→4, 3→0, and 4→0 are detected (in Fig. 5 two detectors are shown). Besides, there is an informational action on the ion, e.g., by modulation of light coming from the source $S$. The modulator $M$ is controlled by the source of information $SI$, which, in turn, is connected with one or more detectors to close the feedback loop.

The feedback works in such a way as to stimulate a channel and rate of transitions, in the case of Fig. 5, the 2→4 and 4→0 transitions. The source $SI$ sends a message, e.g., one line of a page or a measure of a music, only if it receives a signal from detector $D_{24}$. Each next message continues the previous one, i.e., is the next line or the next measure.

If the ion has a consciousness and is interested in the information being proposed, it develops a conditioned reflex and will prefer the 2→4 transition to the 2→3 one. Besides, the rates of both 2→4 and 4→0 transitions must increase. All this can be registered by the experimenter. To be sure that the effect is connected with information, one can make a control experiment to cut off the feedback or/and to use some "trivial" information, etc.

Deviation from the standard transitions may be interpreted as an interest of the ion towards the information. Such an interest is thought to be an inherent attribute of each consciousness. This important result does not even depend on the ability of the ion to decipher information. It is sufficient that it is curious. It is like people of modern times were interested in ancient hieroglyphic symbols long before they learned how to read them.

The sequence of information offered for the ion may be a kind of a course teaching it a language for further dialogue. To measure the progress of learning, the experimenter may sometimes introduce into informational channel some specific "request". For example, one can "ask" the ion to choose 2→3 transition rather than the 2→4 one. Since the ion, eager not to shirk the lessons, will tend to prefer the 2→4 transition, the execution of this request can easily be detected by an experimenter and will mean that the text was decoded.

However, the possibilities of an experiment typified in Fig. 5 are not exhausted by this. Purposefully choosing the transitions, the ion, in its turn, can send information to the experimenter using "right" and "left" (in Fig. 5) transitions as a binary code.

Something more about presented interpretation of QM can be found in [5] and [6].